\documentclass[prb,aps,twocolumn,superscriptaddress]{revtex4-1}
\usepackage{amssymb,amsmath,amsfonts,amsthm}
\usepackage{graphicx}
\usepackage{hyperref}
\usepackage{systeme}
\usepackage[makeroom]{cancel}
\usepackage[utf8]{inputenc}
\usepackage[english]{babel}
\usepackage[usenames, dvipsnames]{color}

\newcommand{\av}[1]{\langle#1\rangle}

\newcommand{\s}[2]{\sigma^{#1}_{#2}} 

\begin{document}

\title{Remnants of Anderson localization in pre-thermalization induced by white noise}

\author{S. Lorenzo}
\affiliation{Dipartimento di Fisica e Chimica, Universit\`a degli Studi di Palermo, Via Archirafi 36, I-90123 Palermo, Italy}  

\author{T. Apollaro}
\affiliation{ICTP East African Institute for Fundamental Research and
University of Rwanda
Kigali, Rwanda}

\author{G. M. Palma}
\affiliation{NEST, Istituto Nanoscienze-CNR and Dipartimento di Fisica e Chimica,
Universit\'{a} degli Studi di Palermo, via Archirafi 36, I-90123 Palermo, Italy}

\author{R. Nandkishore}
\affiliation{Department of Physics and Center for Theory of Quantum Matter, University of Colorado Boulder, Boulder, Colorado 80309, USA}

\author{A. Silva}
\affiliation{SISSA, Via Bonomea 265, I-34136 Trieste, Italy}

\author{J. Marino}
\affiliation{Department of Physics and Center for Theory of Quantum Matter, University of Colorado Boulder, Boulder, Colorado 80309, USA}

\begin{abstract}
We study the non-equilibrium evolution of a one-dimensional    quantum Ising chain with spatially disordered,  time-dependent,  transverse fields characterised by white noise correlation dynamics.
We  establish pre-thermalization in this model, showing that the  quench dynamics of the on-site transverse magnetisation first   approaches  a metastable state unaffected by noise fluctuations, and then relaxes exponentially  fast towards an infinite temperature state as a result of the  noise.  
We also consider energy transport in the model, starting from an inhomogeneous  state with two domain walls which separate regions characterised by spins with opposite transverse magnetization.
We observe  at intermediate time scales a phenomenology akin to Anderson localization: energy remains localized within the two domain walls, until the Markovian  noise destroys coherence and accordingly disorder-induced localization, allowing the system to relax towards the late stages of its non-equilibrium dynamics. 
We compare our results with the simpler case of a noisy quantum Ising chain without disorder, and we find that the pre-thermal plateau  is a generic property of  spin chains with weak noise, while the phenomenon of \emph{pre-thermal Anderson localization} is  a specific feature arising from the competition of noise and disorder in the  real-time transport properties of the system. 
\end{abstract}

\maketitle

\section{Introduction} 
Modern experimental advances in  control of cold atoms~\cite{Bloch2008} have revived the interest in non-equilibrium physics~\cite{Greiner2002a} and in real-time dynamics occurring in isolated quantum systems~\cite{PolkovnikovRMP, Eisert2015a}.
Besides  fundamental questions regarding eventual thermalisation of closed interacting systems, the current interest in out-of-equilibrium physics stands mainly in the possibility to engineer novel phases of matter or in realising phenomena that do not have a  counterpart in traditional statistical mechanics, arising when a quantum many body system is driven far away from equilibrium  for significantly long times.
Noticeable examples range from Floquet topological insulators~\cite{kita, Lindner11} to time crystals~\cite{Monroe, Choi}, encompassing pre-thermalization~\cite{Berges2004a, Gring2012, Langen2016} and dynamical phase transitions~\cite{Zhang}.

{
%
%
%
%
Another prominent example of non-ergodic phase of matter, nowadays accessible with cold gases experiments~\cite{Schreiber2015, Bordia, Choi2016}, is provided by the inhibition of transport in low dimensional disordered systems~\cite{Anderson, AALR} -- a feature persisting even in the presence of many-body interactions~\cite{Nandkishore-2015, VHO, AbaninReview}. 
%
%
While the primary setup to study localization effects in condensed matter platforms are isolated quantum systems, in any practical implementation,   coupling  to the environment is unavoidable, and understanding the interplay of a strongly localized system with  an external equilibrium (or non-equilibrium) bath is of paramount importance, both for   experiments,  as well as   to understand the robustness of Anderson (and many-body) localization to ergodic perturbations.
In this respect, the natural expectation that a bath can facilitate hopping in an otherwise localized system, has been confirmed by  theoretical studies~\cite{NGH, gn, NGADP, BanerjeeAltman, Altmandeph, Levi, Medv16, avalanches, gopknap, deroeck17, Chandran17}  and by a recent cold atoms experiment, where controlled dissipation originates from environmental photons~\cite{Bloch17}.
Despite suggesting the expected fragility of  localized systems to bath-induced decoherence, these works  have also demonstrated    that the interplay of localization and dissipation can imprint interesting signatures on the evolution of physical observables before the eventual onset of relaxation.
%
%
%
}

In this work, we aim at showing how the  characteristic features of an Anderson insulator {--  the inhibition of energy transport across the system -- can persist at intermediate time scales in a simple,  archetypical, disordered quantum spin   system perturbed by  Markovian noise,  by inspecting observables sensitive to transport properties}.
~The dynamics arising after a quantum quench  of isolated, disordered spin models does not tend  towards a steady state in the long time limit~\cite{Ziraldo1, Ziraldo2}, while a quatum Ising model coupled to a Markovian bath via its transverse field  thermalises efficiently with correlations spreading in a light-cone fashion~\cite{Marino2012,Marino2014} (see for instance Ref.~\cite{BanchiPRX17} for further recent developments in this direction).
Here we merge together these two scenarios, considering the quantum quench dynamics of a disordered quantum Ising chain in one dimension (equivalent to a quadratic model of spinless fermions on a lattice), driven by time-dependent noisy transverse fields, and benefiting of these two previously studied  cases~\cite{Marino2012,Marino2014,Ziraldo1, Ziraldo2}  as a benchmark for our results.
Despite  the fact that the localized phase of non-interacting fermions on a lattice is  destroyed by the coupling to a heat bath~\cite{woly, Derrico, NGADP, BanerjeeAltman, Bloch17}, our findings show that on intermediate time scales transport can still be impeded by disorder, and only at longer times quantum coherence is wiped out by the noise, thermalization is established, and energy is free  to redistribute across the system. 

This phenomenon bears analogies with prethermalization in weakly non-integrable systems~\cite{Kollar2011, Bertini2015, Langen2016}, where dynamics is first dominated by  features of the perturbed integrable hamiltonian and only at later times -- when inelastic collisions induced by integrability breaking channels become effective,  the system is capable to relax and dynamics  is attracted by a thermal state.   
Using a similar logic, we first establish the existence of an analogous intermediate regime in our model studying the dynamics of simple observables, as the on-site transverse magnetization, and then we show that remnants of Anderson localization can persist at intermediate times, focusing on specific features of energy transport which become manifest when the Ising chain is  prepared in a spatially inhomogeneous spin state.

\section{A quantum Ising chain \\ with noise and disorder} 
We consider  the real-time dynamics of  the transverse field quantum Ising chain in one dimension~\cite{Sachdevbook} 
\begin{eqnarray}
H_0&=&{-}\sum^n_{j=1}\left(\s{x}{j}\s{x}{j+1}+h\s{z}{j}\right),
\label{H}
\end{eqnarray}
where $\s{x,y,z}{j}$ are  Pauli matrices acting on the  site $j$ of the chain, and $h$ is a uniform transverse magnetic field. This model is characterized by two mutually dual gapped phases separated by a continuous quantum phase transition at $h=1$; it is exactly solvable by a Jordan-Wigner transformation mapping it onto a system of free fermions, which is then diagonalised by a Bogoliubov rotation~\cite{Sachdevbook}. This makes the Ising chain in Eq.~\eqref{H} equivalent to a collection of   free fermions, $\gamma_k$, with momenta 
$k_j=\pm \pi/n(2j+1)$ where $j=0,\dots,\frac{n}{2}-1$.

 \begin{figure} [t!]
 \centering
    \includegraphics[width=8.0cm]{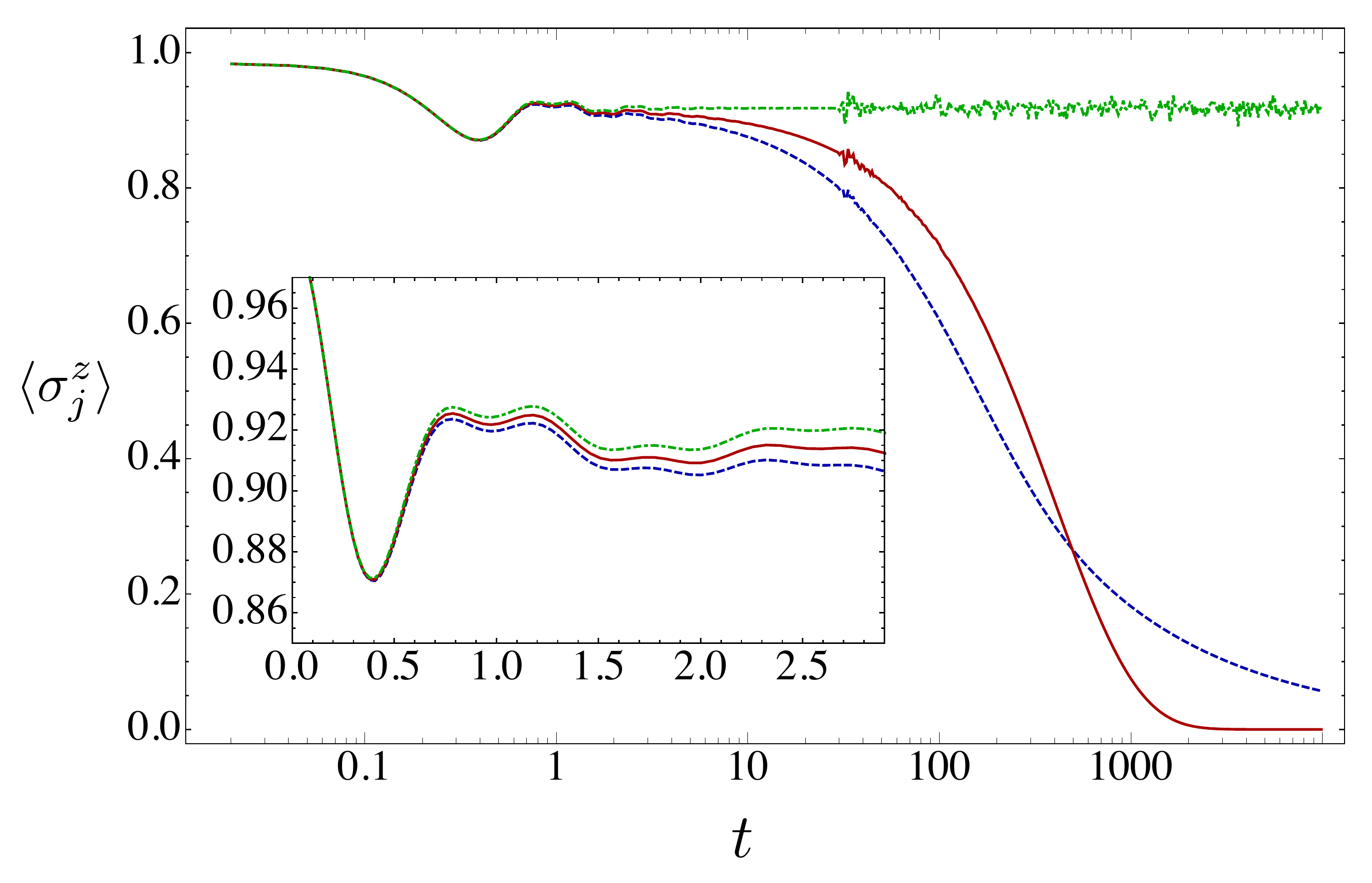}
    \caption{\emph{(Colors online)} Expectation value of the local transverse magnetization, $\langle\sigma^z_j(t)\rangle$, as a function of time, $t$, after a quench from $h_0=4$ to $h=2$ of the transverse field of the quantum Ising model Eq.~\eqref{H}; {here $n=256$.} 
    The green line corresponds to dynamics without noise, and sets the value of the pre-thermal plateau attained under influence of homogeneous (blue line) or inhomogeneous (red line) noise along the transverse field direction ($\Gamma=0.1$). 
    As discussed in the main text and shown in the figure, relaxation towards the  asymptotic equilibrium value occurs faster in the latter case. The inset displays the early stages of the magnetization dynamics.}
     \label{fig1}
\end{figure}
\begin{figure*}
 \centering
    \includegraphics[width=7.0cm]{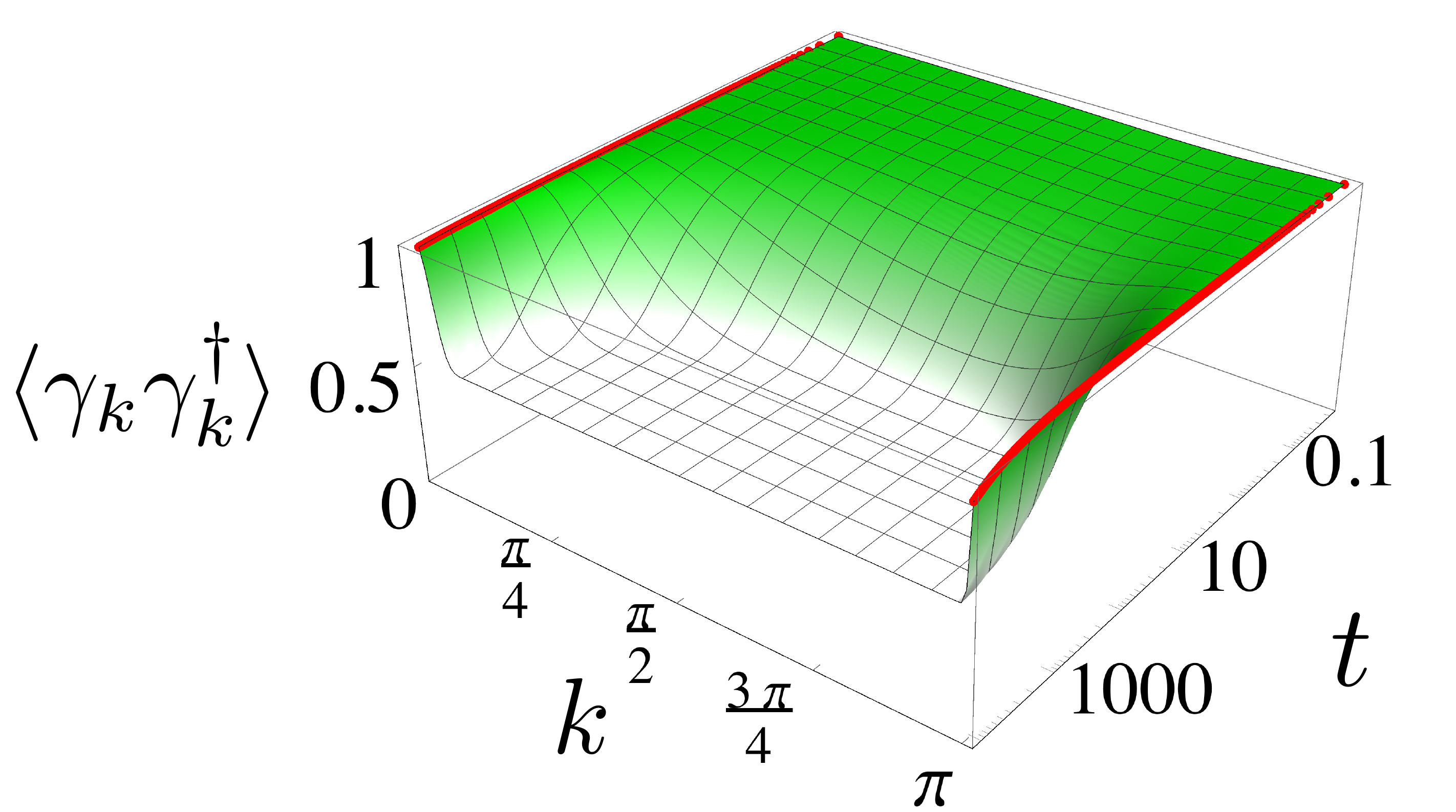}\qquad\qquad\includegraphics[width=7.0cm]{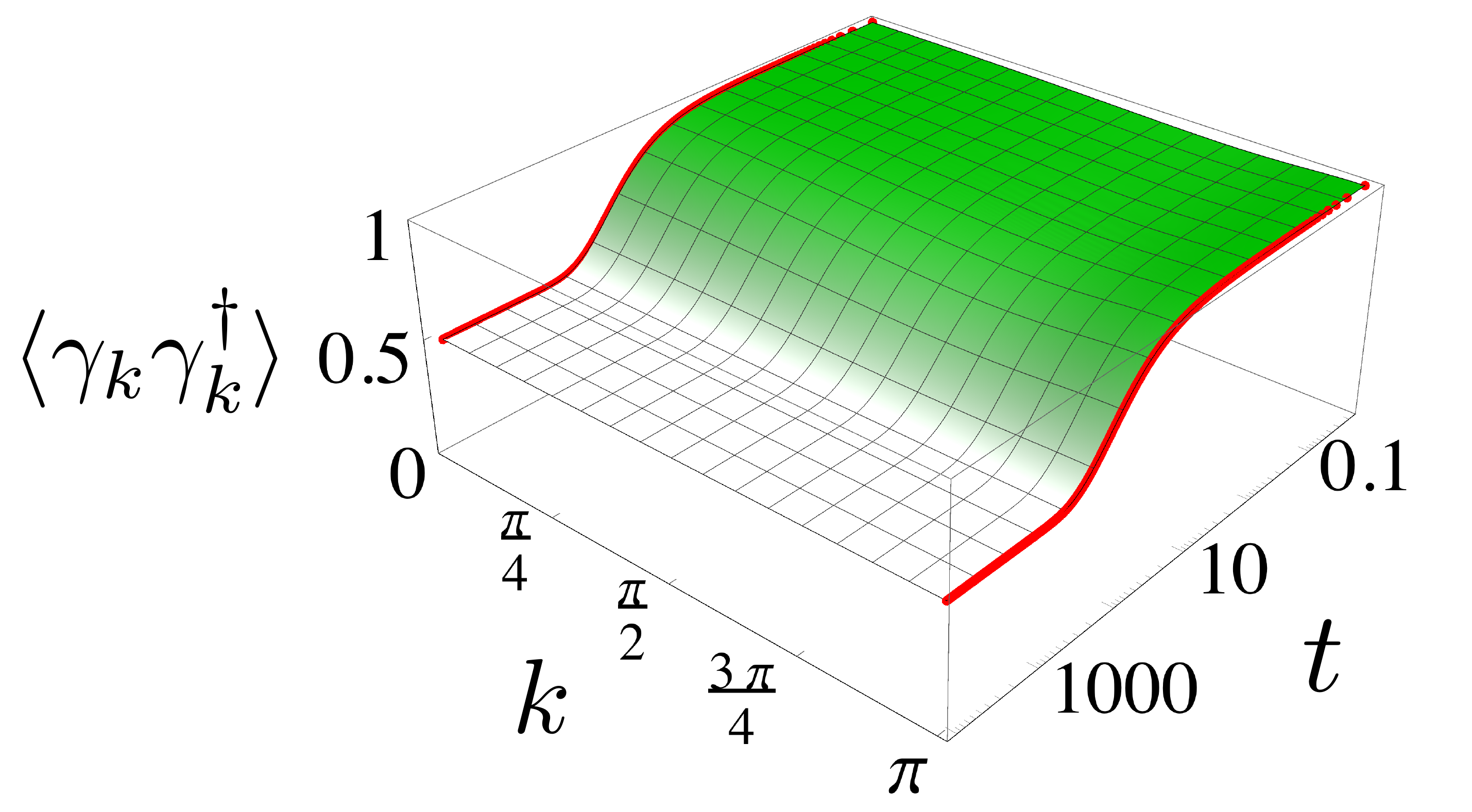}
    \caption{
\emph{(Colors online)} Population, $\langle\gamma_k\gamma^\dag_k\rangle$, of Bogolyubov modes as a function of the momentum, $k$, and time, $t$, after a quantum quench of the transverse field, in the presence of spatially homogeneous (left) and inhomogeneous (right) noise. The modes close to the band edge $k{^*}=0,\pi$  (marked with red continuous lines) undergo a slower evolution under the effect of uniform noise, while, in the instance reported in the right panel,  populations relax for any momentum $k$ swiftly towards their asymptotic infinite temperature value, $\langle\gamma_k\gamma^\dag_k\rangle(t\to\infty)=1/2$. 
Parameters in the plots are the same as in Fig.~\ref{fig1}: $h_0=4$, $h=2$, $\Gamma=0.1$, {$n=256$}.}
     \label{fig2}
\end{figure*}

We will be interested in studying the dynamics of an Ising chain subject to an inhomogeneous time dependent noise. For this sake at $t=0$ we switch on a space and time-dependent Gaussian white noise, $\eta_j(t)$, superimposed to the uniform transverse field, $h$,  on each site $j$ of the chain,  as described by the operator%
\begin{equation}\label{eq:fields}
V(t)=\sum^{n}_{j=1}\eta_j(t)\s{z}{j}.
\end{equation}
The Gaussian field  $\eta_j(t)$ is chosen with zero average $\langle\eta_j(t)\rangle=0$,  and is characterised by the two-point function,
\begin{equation}\label{eq:corrnoise}
\langle\eta_i(t)\eta_j(t')\rangle=\Gamma \delta_{ij}\delta(t-t').
\end{equation}
At a fixed time $t$, $\eta_j(t)$ describes  an inhomogeneous configuration of  transverse fields along the  quantum Ising chain, from site $j=1$ to $j=n$,  drawn from a Gaussian distribution of variance $\Gamma$;
the memoryless nature of  $\eta_j(t)$ ensures that these spatial disorder configurations are generated in an uncorrelated fashion at every time $t$.

We are therefore considering  the non-equilibrium dynamics of  the model
\begin{equation}\label{ham:totale}
H=H_0+V(t),
\end{equation}
describing a quantum Ising chain with competing  time-dependent noise and spatial disorder along the direction of the transverse field.
Equivalently,  Eq.~\eqref{ham:totale} describes disordered, non-interacting  fermions on a one dimensional lattice and driven by a time-dependent Markovian noise.

The evolution of the density matrix of the system, $\tilde{\rho}(t)$, is ruled by the equation  of motion
\begin{eqnarray}
\frac{d}{dt}\tilde{\rho}(t)=-i[H+V(t),\tilde{\rho}(t)].
\label{stoc}
\end{eqnarray}
%
%
%
Following a standard procedure (see for instance Refs.~\cite{Novikov1965, Budini2001, Breuerbook} and the Appendix),   it is possible to derive a local-in-time master equation for the dynamics of the density matrix, $\tilde{\rho}(t)$, averaged over different realisations of the noise field, $\rho(t)\equiv\langle\rho_{\text{st}}(t)\rangle$,
\begin{equation}\label{avME1}
\frac{d}{dt}\rho(t)
{=-i}[H,\rho(t)]{-}\Gamma^2  \sum^n_{j=1}\left[\s{z}{j},\left[\s{z}{j},\rho(t)\right]\right],
\end{equation}
which we solve numerically starting from  different initial non-equilibrium conditions, in order to extract dynamics of observables of interest in this work.
In the following, we will consider both  quantum quenches of the transverse field -- the system is prepared in the ground state of the Hamiltonian~\eqref{H} with a given value of $h_0$ and evolved at later times under the influence of the noise and at a different value of the average transverse field $h$ -- as well as the dynamics starting from spatially inhomogeneous spin states.

Before discussing the results, we recall that the impact of a spatially homogeneous Markovian noise, $\eta(t)$, on the quench dynamics of the quantum Ising chain has been previously studied   by two of us in Refs.~\cite{Marino2012, Marino2014}, using Keldysh diagrammatics methods.
As in the presence of an inhomogeneous field $\eta_i(t)$, an  analogue master equation for $\rho(t)$ can be derived for the homogeneous case, $V_g(t)\propto \eta(t)\sum^n_{j=1}\sigma^z_j$, and  reads
\begin{equation}
\frac{d}{dt}\rho(t)
{=-i}[H,\rho(t)]{-}\Gamma^2 \sum^n_{j,j'=1}\left[\s{z}{j},\left[\s{z}{j'},\rho(t)\right]\right].
\label{avME2}
\end{equation}
The numerical solution of Eq.~\eqref{avME2}, and the analytical results of~\cite{Marino2012, Marino2014}, will be used in the following to benchmark our findings with the non-equilibrium dynamics of the model~\eqref{ham:totale} and its master equation~\eqref{avME1}.

\section{Pre-thermalization \\
induced by Markovian noise} 
We first show that after a quantum quench  $h_0\to h$, the effect of an homogeneous, $\eta(t)$, and an inhomogeneous,  $\eta_i(t)$, noisy transverse field have a  qualitative, similar impact on the dynamics of single-site observables.
In particular, we consider the local transverse magnetisation, $\sigma^z_j$, at a given site $j$, and we calculate numerically the evolution of its expectation value, averaging over the density matrix $\rho(t)$. 
Fig.~\ref{fig1} shows that $\langle\sigma^z_j(t)\rangle$ reaches, after a first relaxation process, a plateau  with an expectation value  close to the one acquired  after a quantum quench  of the Ising chain without noise (if $\Gamma\ll|h-1|$, as in the homogeneous case~\cite{Marino2012, Marino2014}).
This behaviour is  akin to   the phenomenon of pre-thermalization in isolated systems, since it precedes the decay of  $\langle\sigma^z_j(t)\rangle$ towards its actual equilibrium value, which is set by the   infinite temperature state -- since  the Markovian noise, $\eta_j(t)$, can heat  the system indefinitely. 

The runaway of $\langle\sigma^z_j(t)\rangle$ from the pre-thermal state (towards the asymptotic, infinite temperature one) is exponential in time,~{ $\langle\sigma^z_j(t)\rangle\propto e^{-\Sigma t}$, with the rate of decay $\Sigma\propto \Gamma$ (as one can easily check from the numerics)}, in the presence of the inhomogeneous field $\eta_j(t)$, while, {when the Markovian noise is homogeneous}, $\langle\sigma^z_j(t)\rangle$ drops algebraically as $\sim1/\sqrt{t}$ for $t\gg1/\Gamma$ (Fig.~\ref{fig1});
{this latter result was already}  found in Refs.~\cite{Marino2012, Marino2014}, and we have confirmed its validity from the numerical solution of the Lindblad dynamics given by Eq.~\eqref{avME2}.

The different relaxational laws in the two cases are due to the role played by the two modes $k=0,\pi$, which are slow when the Ising chain is driven by an homogeneous noise field and can significantly affect  late-time dynamics.%

For time-dependent  perturbations proportional to the total transverse magnetisation like $V_g(t)$, the occupation number of the two Bogolyubov modes close to the band edges, $k^*=0$ and $k^*=\pi$, are conserved quantities $[n_{k^*},H_0+V_g(t)]=0$, with $n_k=\gamma^\dag_k\gamma_k$. 
This  commutator vanishes continuously when the limits $k\to0$, or $k\to\pi$, are taken,
 implying that the relaxation rates, $\Upsilon_k$, of the modes close to the band edges vanish continuously as well, $\Upsilon_k\propto k^2$, see also Ref.~\cite{Marino2014}, and determining  a slow, algebraic relaxation of one point functions (as the on-site transverse magnetization $\langle\sigma^z_j(t)\rangle$),  which can be expressed as bi-linears  of  Bogolyubov operators, and whose dynamics is accordingly determined by the expectation values  $\langle \gamma_k\gamma^\dag_k(t)\rangle$ -- after coherences, $\langle \gamma^\dag_k\gamma^\dag_{-k}(t)\rangle$, have been suppressed by noise-induced dephasing.
In contrast, for inhomogeneous time-dependent fields as in Eq.~\eqref{eq:fields}, there are no soft  modes slowing down quantum evolution and dissipation drives  quickly the system towards the asymptotic steady state of dynamics.
The two panels of Fig.~\ref{fig2} show a three-dimensional plot of $\langle \gamma_k\gamma^\dag_k\rangle$ as a function of time, $t$, and momentum, $k$, respectively for homogeneous (left panel) and inhomogeneous (right panel) noisy transverse fields.
According to the above discussion on quasi-particles relaxation rates, the figure shows that $\langle \gamma_k\gamma^\dag_k\rangle$ relaxes uniformly for all momenta $k$ in the presence of competing noise and disorder, while in the presence of global noise, the modes with wave-vectors close to $k{^*}=0,\pi$ approach slowly their asymptotic equilibrium value.

\section{Pre-thermal Anderson localization} 
We now extend our study to the energy transport properties of the noisy chain~\eqref{ham:totale}.
First of all, we consider as initial state $|\psi_0\rangle$,  an inhomogeneous spin texture (without performing a quench of the transverse field, $h_0=h$), preparing a  region of spins polarised  along the positive $\hat{z}$ direction  at the centre of the chain,
\begin{widetext}
\begin{equation}\label{eq:inspin}
|\psi_0\rangle=|\downarrow_1\downarrow_2...\downarrow_{n/2-m-1}\uparrow_{n/2-m}\uparrow_{n/2-m+1}...\uparrow_{n/2+m}\downarrow_{n/2+m+1}...\downarrow_{n-1}\downarrow_n\rangle;
\end{equation}
\end{widetext}
 the block of size $2m+1<n$ in the state~\eqref{eq:inspin} is delimited by two domain walls, separating regions with different spin polarizations.
We  let evolve the system under the Lindblad dynamics~\eqref{avME1}, and study the flow of local energy, 
\begin{equation}\label{locene}
h_\ell(t){=}{-}\sigma^x_\ell\sigma^x_{\ell{+}1}{-}\frac{h}{2}(\sigma^z_\ell{+}\sigma^z_{\ell{+}1}),
\end{equation} 
governed by the equation 
\begin{equation}\label{dynloc}
\langle\dot{h}_\ell(t)\rangle
=\langle j_\ell(t)-j_{\ell{+}1}(t)+ \Sigma_\ell(t)\rangle,
\end{equation}
where $j_\ell(t)=h(\sigma^y_\ell\sigma^x_{\ell{-}1}{-}\sigma^y_\ell\sigma^x_{\ell{+}1})$, and $\Sigma_\ell(t)=4\Gamma^2(\sigma^x_\ell\sigma^x_{\ell{+}1})$, and where  the average over the state $\rho(t)$ has been taken.
The term $\Sigma_\ell(t)$ changes into
$
\Sigma_\ell(t)=4\Gamma^2(\sigma^x_\ell\sigma^x_{\ell{+}1}{-}\sigma^y_\ell\sigma^y_{\ell{+}1})
$
when the noise perturbation is homogeneous, $V_g(t)$.
Eq.~\eqref{dynloc} is straightforwardly derived, evolving the local energy~\eqref{locene} with the Lindbladian dynamics encoded in Eq.~\eqref{avME1}.

\begin{figure} [t!]
 \centering
    \includegraphics[width=9cm]{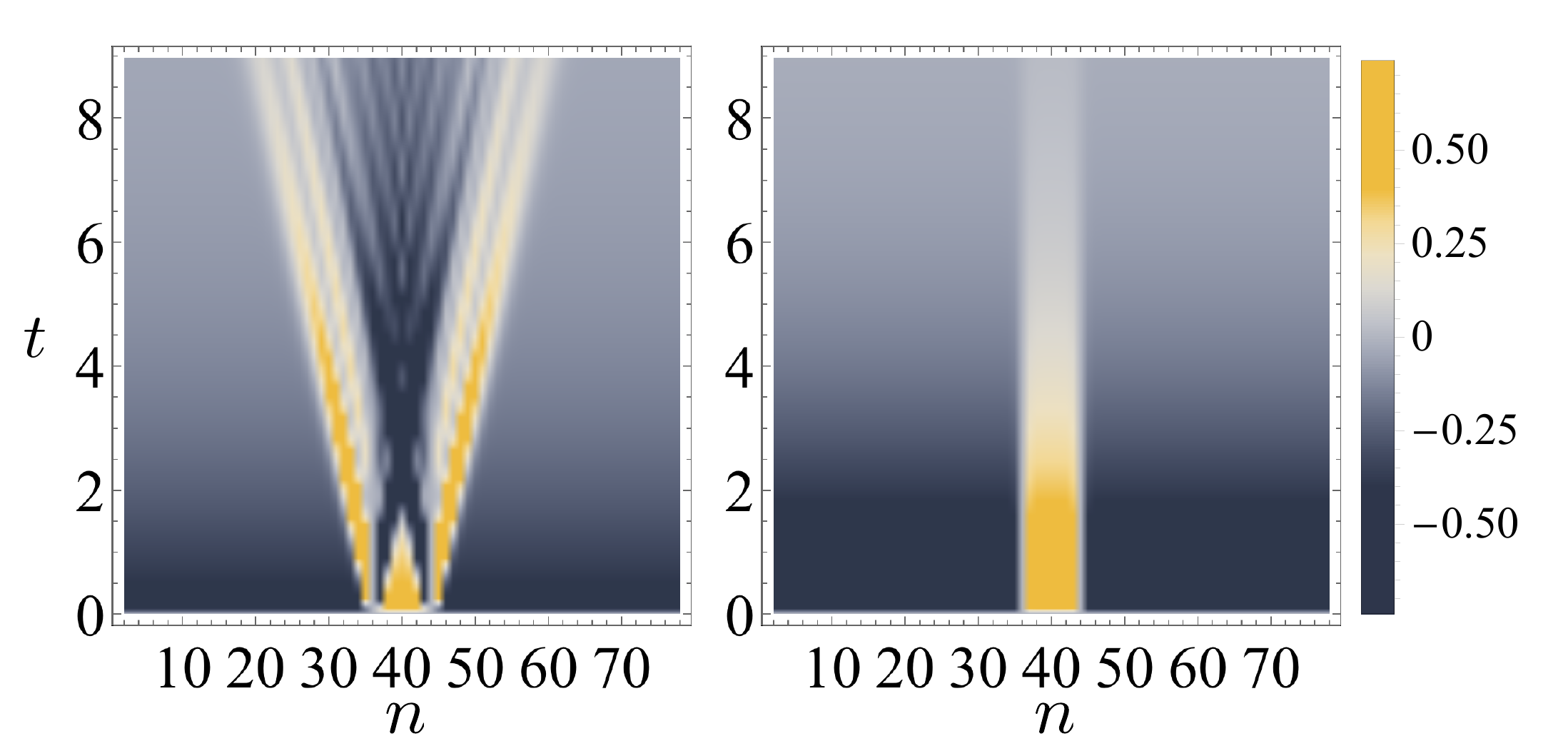}
    \caption{\emph{(Colors online)} Rate of energy flow $\langle\dot{h}_\ell(t)\rangle$ as a function of time, $t$, in a quantum Ising chain of size $n=80$, prepared  with ten spins polarised in the positive $\hat{z}$ direction at its centre ($m=5$).   The left panel corresponds to energy transport in the presence of  homogeneous noise along the transverse field, while in the case of  inhomogeneous noisy transverse fields (right panel), ballistic transport is inhibited by Anderson localization {(in the time window $0<t\lesssim 2$, for the instance of dynamics realised in the figure)}. Blockade of energy transport  is ruled out by  decoherence after a transient (see right panel again), with $\langle\dot{h}_\ell(t)\rangle$ approaching eventually the infinite temperature state. 
    Dynamics of $\langle\dot{h}_\ell(t)\rangle$ has been simulated with $h=2$ and $\Gamma=1.5$ in the figure.}
     \label{flow}
\end{figure}

Fig.~\ref{flow} shows time evolution of the expectation value of the rate of energy flow $\langle\dot{h}_\ell(t)\rangle$, at every site $l$, starting from an inhomogeneous spin state of the type~\eqref{eq:inspin} with $m=5$, in a chain of length $n=80$. 

The left side of Fig.~\ref{flow} corresponds to evolution under the collective homogeneous noise field, $V_g(t)$, while the right side shows dynamics  driven by the inhomogeneous one. 
The difference among the two is noticeable. 
In the first case, a linear light-cone propagation rules the transport of the energy, initially stored in the region of size $2m+1$ at the center of the chain, towards the borders; this finding is consistent with the light cone structure  of  spin correlation functions in a quantum Ising model driven by global noise, $V_g(t)$, found in Refs.~\cite{Marino2012, Marino2014}.

The most striking effect is demonstrated in the second panel of Fig.~\ref{flow}.
The fields $\eta_i(t)$ act equally as spatial disorder and Markovian noise (see discussion after Eq.~\eqref{eq:corrnoise}), and  they compete in order to determine the transport properties of the model~\eqref{ham:totale}.
Energy transport is inhibited at short times:  a disordered non-interacting model undergoes Anderson localization at any disorder strength in one dimension~\cite{Anderson, Anderson2} and this is reflected in the trapping of energy within the region  of size $2m+1$ at the center of the Ising chain (cfr. with right panel of Fig.~\ref{flow}). 
However, since disorder-induced localization originates from quantum interference among wave packets scattering against  disordered lattice centres (represented by the fields $\eta_j(t)$ on the  sites $j$), the blockade of energy transport will  persist until decoherence induced by the Markovian becomes sizeable. 
At that point, quantum coherence is  washed out, Anderson localization disappears,  and energy is left free to spread. 
However, at comparable times, $\langle\dot{h}_l(t)\rangle$ will approach the trivial infinite temperature state, as it occurs  in the dynamics of the on-site transverse magnetization, $\langle\sigma^z_l(t)\rangle$.
The effect is prominent for a disorder variance, $\Gamma$, comparable to the transverse field, $h$; 
for smaller values of $\Gamma$, energy transport would become sizeable again, consistently with previous studies reporting  diffusion in noisy Anderson models~\cite{woly, Derrico}. 
{However, $\Gamma$ cannot be excessively large since it also controls the time scales for the onset of decoherence  and, accordingly, for the  disappearance of energy localisation effects (from numerics we observe that this occurs at times of the order of $1/\Gamma$). Therefore, the inhibition of transport occurring at intermediate times and reminiscent of Anderson localization requires a  variance, $\Gamma$, sufficiently large to start with a localised state, but at the same time tuned to make the effect visible   for an appreciable time window.}

The  phenomenon above can be described as a \emph{pre-thermal Anderson localization}: despite the fact that a disordered  system coupled to an infinite temperature bath cannot display a localized phase, the confining effect of disorder is active at intermediate time scales {(for the instance of dynamics realised in the right panel of Fig.~3, this occurs in the time window $0<t\lesssim 2$)}. This is reminiscent of pre-thermalization in non-integrable closed systems, where features of the weakly perturbed integrable dynamics can persist at intermediate times before eventual equilibration (ruled by integrability breaking perturbations) occurs~\cite{Kollar2011, Bertini2015, Langen2016}. 

{We remark though that the energy flux $\dot{h}_l(t)$ wouldn't display any appreciable evolution in a one dimensional, disordered, quantum Ising chain without noise, since energy transport would be inhibited by Anderson localization. This explains the unusual pre-thermal dynamics of $\dot{h}_l(t)$, compared, for instance, to the one in Fig.~\ref{fig1}. Specifically, in the spirit of pre-thermalization, the dynamics of $\dot{h}_l(t)$ is first ruled by the disordered spin chain without noise (displaying accordingly no significant evolution), while at later times the thermalising effect of the noise becomes significant and induces relaxation to a fully mixed state.}  

{This atypical form of pre-thermal dynamics, where a first relaxational process is absent, is therefore a specific feature of the observable employed to monitor energy spreading in the Ising chain, and a consequence of lack of transport dynamics in a one dimensional Anderson insulator. On the other hand, the dynamical features displayed in the evolution of  one-point observables, as $\langle \sigma^z_j\rangle(t)$ (Fig.~1), would be insufficient to reach conclusive statements on the presence of Anderson localisation in the first relaxational plateau of a system coupled to a noisy environment like the one discussed in this work. 
A possible extension to demonstrate a cleaner two step relaxation in the dynamics of energy transport, could consist in considering a disordered quantum spin chain supporting a many-body localisation transition. Deep in the MBL phase, each spin can be expanded in the basis of local integrals of motion, and will undergo a non-trivial dynamics, exhibiting in general a long-time equilibrated expectation value (as discussed for instance in~\cite{An15}). Extending a similar analysis to the study of transport sensitive quantities, like $\dot{h}_l(t)$, could further substantiate our claim of remnants of disorder induced localisation effects in a two-step dynamical relaxation process. }

%
\section{Perspectives} 
An interesting extension of our result would consist in studying the robustness of the pre-thermal Anderson phenomenon to  integrability breaking  perturbations of  the Ising hamiltonian~\eqref{H} (e.g. a next-neighbour spin-spin interaction in the transverse direction, $U\propto\sum_i\sigma^z_i\sigma^z_{i+1}$), in the spirit of the many-body localization problem (MBL)~\cite{Nandkishore-2015, VHO, AbaninReview}. 
Since the many body localized phase shares, at strong disorder, some features of a genuine Anderson insulator~\cite{HNO, Serbyn, Imbrie, Imbriereview}, 
we expect a qualitative similar phenomenon as the one reported in Fig.~\ref{flow} to manifest (see Refs.~\cite{Altman,Levi} for related studies).
We believe however that generalising a transient MBL behaviour to more complex spin chains (XXZ spin chain) or to different microscopic degrees of freedom (disordered Bose-Hubbard model), has  the potential to highlight a richer phenomenology compared to the one established in this work.
On short/intermediate time scales, where pre-thermal effects set in,  this kind of extensions should be accessible with state-of-art numerical methods.

%
%
%

\section{Acknowledgements } This material is based upon work supported by the Air Force Office of Scientific Research under award
number FA9550-17-1-0183.
S.L, T.A., and G.M.P., acknowledge support from the EU Collaborative Project QuProCS (Grant Agreement No. 641277). \\

\section{Appendix: Derivation of Eq.~(6)}

We consider the average, over different realizations of the noise, of the stochastic unitary dynamics (Eq.(5) in the main text) 
\begin{eqnarray}
\frac{d}{dt}\av{\tilde{\rho}(t)}=-i\av{[H+V(t),\tilde{\rho}(t)]}.
\label{stocAPP}\end{eqnarray}
The result in Ref.~\cite{Novikov1965} gives an exact result for the following mean values, provided the noise is Gaussian:
\begin{eqnarray}
\!\!\!\!\!\!\!\!\!\!\!\!&\langle V(t)\tilde\rho(t)\rangle=\sum_{ij}\int_0^t dt_1 \chi_{ij}(t,t_1)&\;\;\s{z}{i}\Big\langle\frac{\delta\tilde\rho(t)}{\delta\eta_j(t_1)}\Big\rangle,\nonumber\\
\!\!\!\!\!\!\!\!\!\!\!\!&\langle\tilde\rho(t) V(t)\rangle=\sum_{ij}\int_0^t dt_1 \chi_{ij}(t,t_1)&\;\;\Big\langle\frac{\delta\tilde\rho(t)}{\delta\eta_j(t_1)}\Big\rangle\s{z}{i},
\end{eqnarray}
where $\chi_{ij}(t,t')$ is the two point correlation function of the noise resolved in time and space, and where we have used the hermiticity of  $V(t)$.
Substituting the latter results in Eq.~\eqref{stocAPP}, we find
\begin{equation}
\begin{split}
&\frac{d}{dt}\langle\tilde\rho(t)\rangle=
-i[H,\langle\tilde\rho(t)\rangle]+\\
&-i\Delta h \sum_{ij}\int_0^t dt_1 \chi_{ij}(t,t_1)\Big[\s{z}{i},\Big\langle\frac{\delta\tilde\rho(t)}{\delta\eta_j(t_1)}\Big\rangle\Big]. 
\label{avME}\end{split}\end{equation}
We now need to evaluate the response function $\delta\tilde\rho(t)/\delta\eta_{j}(t_1)$ occurring in Eq.~\eqref{avME}.
We assume that at the initial time, the system and the noises are uncorrelated, 
and following Ref.~\cite{Budini2001}, we first formally integrate Eq.~\eqref{stocAPP} in time, and then take a functional derivative with respect to $\eta_j(t_1)$ and $t$, finding %
\begin{eqnarray}
\frac{d}{dt}\frac{\delta\tilde\rho(t)}{\delta\eta_j(t_1)}=-i\Big[H+V(t),\frac{\delta\tilde\rho(t)}{\delta\eta_j(t_1)}\Big].
\end{eqnarray}
The variational derivative satisfy the same Liouvillian equation of the stochastic density matrix, $\delta\tilde\rho(t)$, therefore we can write 
\begin{equation}
\frac{\delta\tilde\rho(t)}{\delta\eta_j(t_1)}=-i\Gamma G(t,t_1)[\s{z}{j},\tilde\rho(t_1)]G^\dagger(t,t_1),
\label{deltarho}\end{equation}
where $G(t,t_1)=\mathcal{T}e^{-i\int_{t_1}^t d\tau(H+V(\tau))}$.

Rewriting Eq.~\eqref{deltarho} as
\begin{equation}
\frac{\delta\tilde\rho(t)}{\delta\eta_j(t_1)}=-i\Gamma [G(t,t_1)\s{z}{j}G^\dagger(t,t_1),\tilde\rho(t)],
\end{equation}
and substituting in Eq.~\eqref{avME} we end up with 
\begin{equation}
\begin{split}
&\frac{d}{dt}\langle\tilde\rho(t)\rangle=
-i[H,\langle\tilde\rho(t)\rangle]-\\
&+\Gamma^2 \sum_{ij}\int_0^t dt_1 \chi_{ij}(t,t_1)\times \nonumber
\Big[\s{z}{i},\Big\langle[G(t,t_1)\s{z}{j}G^\dagger(t,t_1),\tilde\rho(t)]\Big\rangle\Big]. 
\label{avME2}
\end{split}\end{equation}
which for the kind of delta-correlated Markovian noise considered in our work, see Eq. (3) in  the main text, yields directly Eq. (6) of the main text, which we use to simulate the non-equilibrium dynamics of the model.

\bibliography{biblio}

\end{document}